\documentclass{INTERSPEECH2023}
\interspeechcameraready

\title{A Snoring Sound Dataset for Body Position Recognition: Collection, Annotation, and Analysis}
\name{Li Xiao$^1$, Xiuping Yang$^{2,\dagger}$, Xinhong Li$^1$, Weiping Tu$^{1,}$\textsuperscript{\Letter}, \\Xiong Chen$^{2,}$\textsuperscript{\Letter}, Weiyan Yi$^1$, Jie Lin$^1$, Yuhong Yang$^1$, Yanzhen Ren$^3$\thanks{$\dagger$ Equal contribution}\thanks{\Letter Corresponding author}}

\address{
  $^1$NERCMS, School of Computer Science, Wuhan University\\
  $^2$Sleep Medicine Centre, Zhongnan Hospital of Wuhan University\\
  $^3$School of Cyber Science and Engineering, Wuhan University}
\email{tuweiping@whu.edu.cn, zn\_chenxiong@whu.edu.cn}

\begin{document}

\maketitle

\begin{abstract}


Obstructive Sleep Apnea-Hypopnea Syndrome (OSAHS) is a chronic breathing disorder caused by a blockage in the upper airways. Snoring is a prominent symptom of OSAHS, and previous studies have attempted to identify the obstruction site of the upper airways by snoring sounds. Despite some progress, the classification of the obstruction site remains challenging in real-world clinical settings due to the influence of sleep body position on upper airways. To address this challenge, this paper proposes a snore-based sleep body position recognition dataset (SSBPR) consisting of 7570 snoring recordings, which comprises six distinct labels for sleep body position: supine, supine but left lateral head, supine but right lateral head, left-side lying, right-side lying and prone. Experimental results show that snoring sounds exhibit certain acoustic features that enable their effective utilization for identifying body posture during sleep in real-world scenarios.

\end{abstract}

\section{Introduction}

Obstructive sleep apnea-hypopnea syndrome (OSAHS) is a severe chronic breathing disorder and is caused due to a blockage or collapse of the upper airways~\cite{osa,diagnosis1}. The gold standard approach for diagnosing OSAHS is attended overnight polysomnography (PSG) in a sleep laboratory~\cite{punjabi2008epidemiology,dempsey2010pathophysiology}. Effective management of moderate and severe patients with OSAHS requires accurate identification of the site of upper airway obstruction. The standard diagnostic approach to determine the obstructive site is through the Drug-Induced Sleep Endoscopy (DISE) procedure~\cite{dise1}. This tool assesses the upper airway of OSAHS patients while simulating the conditions of natural sleep. However, this method of diagnosis is associated with several limitations, such as extended examination time and elevated costs.

Snoring is one of the most prominent symptoms of OSAHS and can be used to identify the obstructive site, including the velum, oropharyngeal lateral walls, tongue, and epiglottis~\cite{mpssc}. Numerous studies have been conducted in this field by Munich-Passau snore sound corpus (MPSSC)~\cite{mpssc}, with a focus on the automatic classification of snoring sound excitation location~\cite{mpssc-ref4,snore-gan,vote1,vote2}. For example,  Zhang et al.\cite{snore-gan} proposed a novel data augmentation approach utilizing semi-supervised conditional generative adversarial networks. Their study demonstrates significant competitiveness compared to previous work on the same dataset, achieving an unweighted average recall (UAR) of 67.4\%. Ding et al.~\cite{vote2} demonstrated the efficacy of a convolutional neural network and complement-cross-entropy loss function in solving the problem of imbalance distribution of the dataset, yielding a UAR of 77.13\%.

Nevertheless, automatic snoring sound excitation location classification remains challenging in a real-world clinical setting mainly because of changes in the sleep positions of patients. The patients usually lie down supine on the operating table during DISE~\cite{dise2},  and the snoring sounds of MPSSC datasets are classified based on the simultaneous endoscopic video recordings of DISE. However,  sleep body position could influence the site, direction, and severity of upper airway obstruction in patients with OSAHS~\cite{head1,head2}. In the natural sleep environment, people change their sleeping position several times nightly, affecting the corresponding snoring sounds generated and their excitation location in the upper airway.  Therefore, an important relationship exists between sleep posture and the snoring sounds generated. The lack of information on sleep body position could become one of the significant barriers to influencing the result of predicting obstruction sites in the upper airway only based on snoring sounds.  Most of the existing research about sleep posture recognition relies upon using camera or pressure sensors~\cite{sensor1,sensor2,sensor3}. However,  they engender apprehensions about privacy and impose a financial burden associated with the acquisition and maintenance of hardware.

In this paper, a novel snore-based sleep body position recognition dataset (SSBPR) is presented, and the snoring sounds of SSBPR dataset are annotated based on simultaneous PSG signals. The aim of this study is to enhance the accuracy of predicting the site of upper airway obstruction in a clinical setting by body position information only derived from snoring sounds. Specifically, a two-stage method could be used: 1) sleep position recognition by SSBPR  and 2) obstructive site identification by other datasets, such as MPSSC. Therefore, human sleep body positions which influence the upper airway are considered as much as possible in this paper, including supine, supine but left lateral head, supine but right lateral head, left-side lying, right-side lying and prone. Furthermore, motivated by the recent advancements in attention-based models in the field of computer vision and natural language processing~\cite{swin-transformer,chatgpt}, the present study employs AST~\cite{ast}, a model based on Transformer, as the baseline method for sleep body position classification.  Experimental results show that our work holds significant potential in terms of offering practical solutions that could enhance the diagnosis of OSAHS and the management of sleep posture for special populations.

To be more specific, the key contributions of our work are as follows:
\begin{enumerate}[noitemsep,label=\arabic*)]

     \item We construct a snoring sound dataset named SSBPR for sleep body position recognition. The dataset can serve as a valuable supplementary resource to the  MPSSC and future snoring dataset of upper airway obstruction, which can improve the precision and reliability of identifying the location of upper airway obstruction.

    \item The SSBPR dataset has potential applications beyond helping to identify upper airway obstruction sites. Specifically, SSBPR can aid in the monitoring and management of sleep body position in specific populations with snoring, such as pregnant women and patients with laryngopharyngeal reflux disease.

\end{enumerate}

In the remainder of this paper, we first describe our dataset in more detail in section \ref{dataset}, then discuss our baseline experimental results and finally provide conclusions in section \ref{experiments} and \ref{conclusion}.

\section{Datasets}
\label{dataset}
The purpose of the SSBPR dataset is to provide a new benchmark and further advance the research on the automatic snoring sound excitation location classification. In this section, we describe the details of the SSBPR dataset.

\subsection{Data Collection}

In this study, data were collected from 20 adult patients who underwent overnight PSG at a local Sleep Medicine Research Center within the hospital. The study was conducted with the approval of the local medical ethics committee, and patients provided signed consent for their participation, including audio and video recordings during sleep. The personal information of the study subjects was collected and stored anonymously to ensure privacy protection.

\begin{figure}[tp]
    \centering
       \includegraphics[width=0.6\columnwidth]{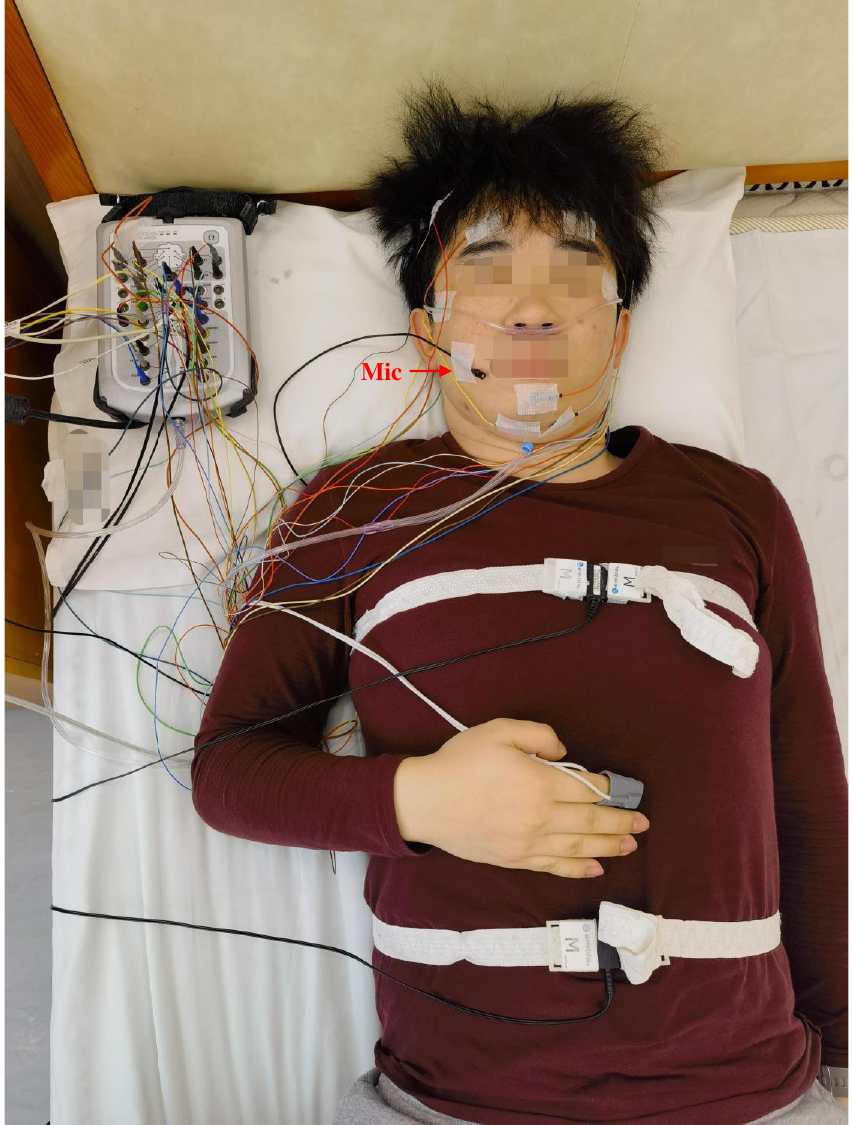}
    \caption{Diagram of the patient's face frame of PSG and location of the microphone in the sleep medicine research center. Sensors attached to the patient's face and body are part of the PSG acquisition system.}
    \label{record-position}
\end{figure}

The snoring audio recording was obtained using a subminiature lavalier microphone (Shure MX150/C, USA) driven by an audio Interface (Antelope Zen Go Synergy Core, Bulgaria), with a sampling frequency of 32 kHz and a sampling resolution of 16 bits (bit depth). The microphone was positioned on the patient's face, facing the mouth, with an approximate distance of 3 cm, as shown in Figure~\ref{record-position}. The microphone was secured in place using medical tape, ensuring stability during the entire night. Recording snoring sounds at a distance of 3 cm offers a range of benefits in terms of capturing precise and detailed information about the sound. The proximity of the recording device enables it to capture subtle nuances in the snoring sounds, such as variations in volume and pitch, which may not be evident when recording at greater distances. This high level of detail can be beneficial for analyzing and diagnosing health issues such as sleep apnea. In addition, recording at a close distance reduces background noise, like traffic or other sounds in the house, making it result in a recording of the snoring sound that is more focused and clearer.

\subsection{Data Annotation}
A standard full PSG simultaneously records more than 10 physiologic signals during sleep, including electroencephalogram (EEG), electrooculogram (EOG),  electromyogram (EMG),  body position, video recording and et al. For recorded PSG data of each patient in the SSBPR dataset, sleep stages were scored by three experienced experts according to the AASM Manual V2.6~\cite{aasm}. A certified technician performed the first level of scoring, and the final level was conducted by two certified doctors who verified the scoring of PSG. Both of them got RPSGT (Registered Polysomnographic Technologist) certified. Compumedics Profusion Sleeps was used to record the data, display multiple channel recordings, manually score the data, report and export the recorded data.

In half of OSAHS patients, disease severity depends on trunk position and head position, in which the supine head and trunk position is usually the worst sleeping position~\cite{position,position2}. In positional OSAHS patients, lateral head rotation alone significantly differed at all levels observed during DISE compared to lateral head and trunk rotation~\cite{head1,head2}. The snoring sounds are usually influenced by changes in the structure of the upper airway. These findings underscore the importance of considering both trunk and head position in accurately monitoring the upper airway of OSAHS patients. Therefore, to effectively monitor the upper airway of patients with OSAHS, it is necessary to label snoring sounds using trunk and head position information.

\renewcommand{\dblfloatpagefraction}{.9}
\begin{figure*}[htp]
    \centering
    \includegraphics[width=1.85\columnwidth]{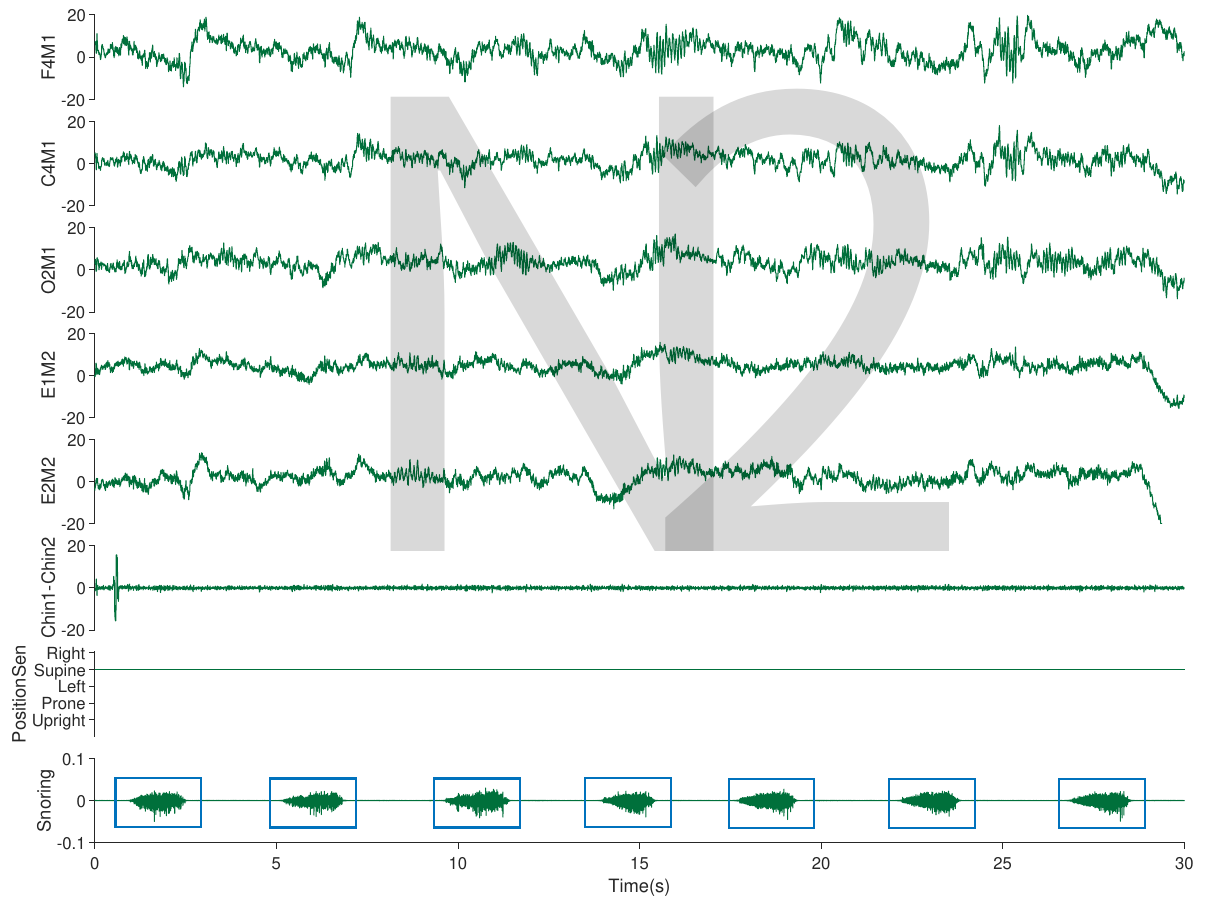}
    
    \caption{An example of snoring data annotation using PSG signals. EEG: F4M1, C4M1 and O2M1; EOG: E1M2 and E2M2; EMG: Chin1-Chin2;  Trunk position: PositionSen. EEG, EOG and EMG are employed to determine the sleep state of the individual, while a body position sensor is utilized to ascertain the positioning of the trunk. Additionally, video monitoring is employed to evaluate the position of the patient's head (In order to protect the patient's privacy, the video signals are not given in this paper). Blue boxes indicate the snoring sounds during sleep while the patient is supine, and the watermark "N2" in the figure indicates that the patient is in non-rapid eye movement stage N2.}
    \label{psg-snoring}
\end{figure*}

In this paper, the annotation of snoring data (duration of 0.29 - 8.39 s) in natural sleep requires synchronized PSG signals: sleep stages, body position and video. The human sleep stages include wake, non-rapid eye movement (NREM) and rapid eye movement (REM) sleep, where NREM sleep encompasses three sleep stages, referred to as stage 1 (N1), stage 2 (N2), and stage 3 (N3) NREM sleep~\cite{sleep-stage}. As shown in Figure \ref{psg-snoring}, the watermark "N2"  indicates that the patient is in non-rapid eye movement stage N2 by EEG, EOG and EMG, which is used to determine whether the patient is asleep. The position and video signal together to identify the trunk and head position of the patient when snoring. In summary, the snoring sounds in the SSBPR dataset are labeled with 6 types:  supine, supine but left lateral head, supine but right lateral head, left-side lying, right-side lying and prone.

\begin{table}[t]
\centering
\caption{The number of classed instances in each set of the SSBPR. The labels 0, 1, 2, 3, 4 and 5 indicate different body positions: supine, supine but left lateral head, supine but right lateral head, left-side lying, right-side lying and prone, respectively.}
\setlength{\tabcolsep}{1mm}{
\begin{tabular}{cccccccc}
\toprule[1pt]   
\multirow{2}{*}{Labels} & \multicolumn{2}{c}{Train} & \multicolumn{2}{c}{Validation} & \multicolumn{2}{c}{Test} & \multirow{2}{*}{$\sum$} \\ \cline{2-7}
                        & Male       & Female       & Male      & Female      & Male       & Female      &                      \\ \hline
0                       & 491        & 491          & 164       & 164          & 164       & 164          & 1638                 \\
1                       & 436        & 436          & 146       &146          & 145        & 145         & 1454                 \\
2                       & 381        & 381          & 127      & 127          & 127        & 127          & 1270                 \\
3                       & 500        & 500          & 167       & 167          & 167        & 167          & 1668                 \\
4                       & 431        & 431          & 144       & 144          & 144        & 144          & 1438                 \\
5                       & 61         & 0            & 20        & 0           & 21         & 0           & 102                  \\ \hline
$\sum$        & 2300       & 2239       & 768     & 748       & 768       & 747        &7570                 \\  \bottomrule[1pt]
\end{tabular}
}
\label{numbers}
\end{table}

\subsection{Data Distribution}

Research studies have consistently demonstrated that OSAHS is more prevalent among men compared to women~\cite{male-female}. To ensure equitable representation of genders in the study sample, a deliberate effort was made to include an equal number of male and female patients in the dataset, with a total of 10 patients of each gender being selected. Given that age and obesity are recognized risk factors for OSAHS~\cite{age}, older individuals and those with higher body mass index may exhibit a more collapsible airway during sleep. Consistent with this notion, the sample population comprises 20 individuals aged 26 to 57 years, with a mean age of 43.1. The patients in this study exhibit a mean body mass index (BMI) of 26.57 kg/$m^2$. The SSBPR dataset contains a total of 7570 snoring samples. Both genders had equal snoring sound samples except for prone labeling because female patients did not appear prone during PSG monitoring. Detailed data split information can be found in Table \ref{numbers}.

\section{Baseline Experiments}
\label{experiments}

\subsection{Baseline Method}

\label{method}

The Transformer architecture has gained widespread usage in many areas, including natural language processing, image processing and audio processing~\cite{swin-transformer,chatgpt,ast}. This popularity can be attributed to the Transformer's capability to effectively model long-term dependencies, which are crucial in numerous tasks that involve sequences. Snoring signals possess complex dependencies between various temporal steps due to changes in the upper airway. The self-attention mechanism of the Transformer facilitates the weighing of the contribution of each temporal step in predicting the current step, thereby capturing the inter-temporal relationships effectively of snoring sounds.

For baseline experiments, we use an AST~\cite{ast}  audio classifier, which has the same Transformer architecture as Vision Transformer (ViT)~\cite{vison1}. AST is the first convolution-free, a purely attention-based model for audio classification, which supports variable length input and can be applied to various tasks. In this paper, AST partitions the 2D snoring sound spectrogram into a sequence of 16 $\times$ 16 patches with overlapping areas. Each patch is then linearly transformed into a sequence of 1-D patch embeddings, augmented with learnable positional embeddings. The resulting sequence is then prepended with a classification token before input into a Transformer. The classification token output is obtained by applying a linear layer to the output embedding, which is used for the final classification task.

\subsection{Experiment Settings }

\textbf{Model inputs.}
Snoring sound and speech exhibit significant acoustic similarities in their generation and emission~\cite{snoring-gene}. Both are produced in the vocal tract through airflow vibration and are acoustically shaped by the frequency transfer function of the upper airway, before being emitted through the mouth and nose. Given that certain acoustic features have proven to be effective in speech-related machine learning tasks, it is plausible that they would be equally applicable to the analysis of snoring. With this in mind, we follow the AST~\cite{ast} to extract log Mel spectrograms, with a frequency dimension of 128, using a 25 ms Hamming window and a hop length of 10 ms. This gives us a resultant input size of 128 × 100t for t seconds of audio.

\textbf{Baseline experiments.}
The present study involved the implementation of the following  baseline experiments using the proposed (SSBPR) dataset. 
\begin{enumerate}[noitemsep,label=\arabic*)]

     \item Since gender is a significant OSAHS risk factor and female patients in our dataset did not snore in a prone position, we conduct a 5-class classification experiment on the SSBPR dataset. We aim to prove the ability of the model to identify sleep positions based on snoring for different genders.
     
     \item We evaluate the various sleep positions of the SSBPR by a six-classes classification experiment on the SSBPR dataset. This experiment aims to assess the ability of the dataset to accurately classify sleep body positions among patients with OSAHS and special patient populations (pregnant women and patients with laryngopharyngeal reflux disease).

\end{enumerate}

\begin{table}[t]
\centering
\caption{Five-classes and six-classes sleep body position classification results.}
\setlength{\tabcolsep}{5mm}{
\begin{tabular}{ccc}
\toprule[1pt]
\#                            & Gender & Accuracy (\%) \\ \hline
\multirow{2}{*}{Five-classes} & Male   & 82.7              \\
                              & Female & 94.6             \\
Six-classes                   & /      & 85.8     \\  \bottomrule[1pt]
\end{tabular}
}
\label{class-acc}
\end{table}

\textbf{Evaluation metric.}
The aim of the SSBPR dataset is to identify the sleeping posture of patients during snoring, so we employ \emph{Accuracy} as the main metric to evaluate the performance of method.

\textbf{Training details.}
The model has been trained on two NVIDIA GeForce RTX 3090 GPUs with batch size 12 for 30 epochs. The experimental setup mostly follows the AST. The model utilizes the initial learning rate of 1e-5 with Adam optimization~\cite{adam}  and binary cross-entropy loss. The warm-up step is set to 5 epochs and the learning rate is scheduled with a factor of 0.85 for every epoch. We train both models by using frequency/time masking data augmentation~\cite{augmention-hz} with max time mask length of 48 frames and max frequency mask length of 48 bins.

\subsection{Results}

As shown in Table \ref{class-acc}, the model achieves an accuracy of 82.7\%, 94.6\% and 85.8\% on male, female subjects and all subjects, respectively. The results show that SSBPR can be used to identify the specific body position of patients and help to diagnose the obstruction site in the upper airway accurately. However, the accuracy for identifying sleeping positions is higher for female patients than for male patients in the five-classes classification experiment.  Our inference is that one of the reasons is some pathophysiological differences between men and women~\cite{gender-difference}. Therefore, the accuracy of the methods can be improved by developing specific algorithms for male patients.

\begin{figure}[tp]
    \centering
       \includegraphics[width=0.8\columnwidth]{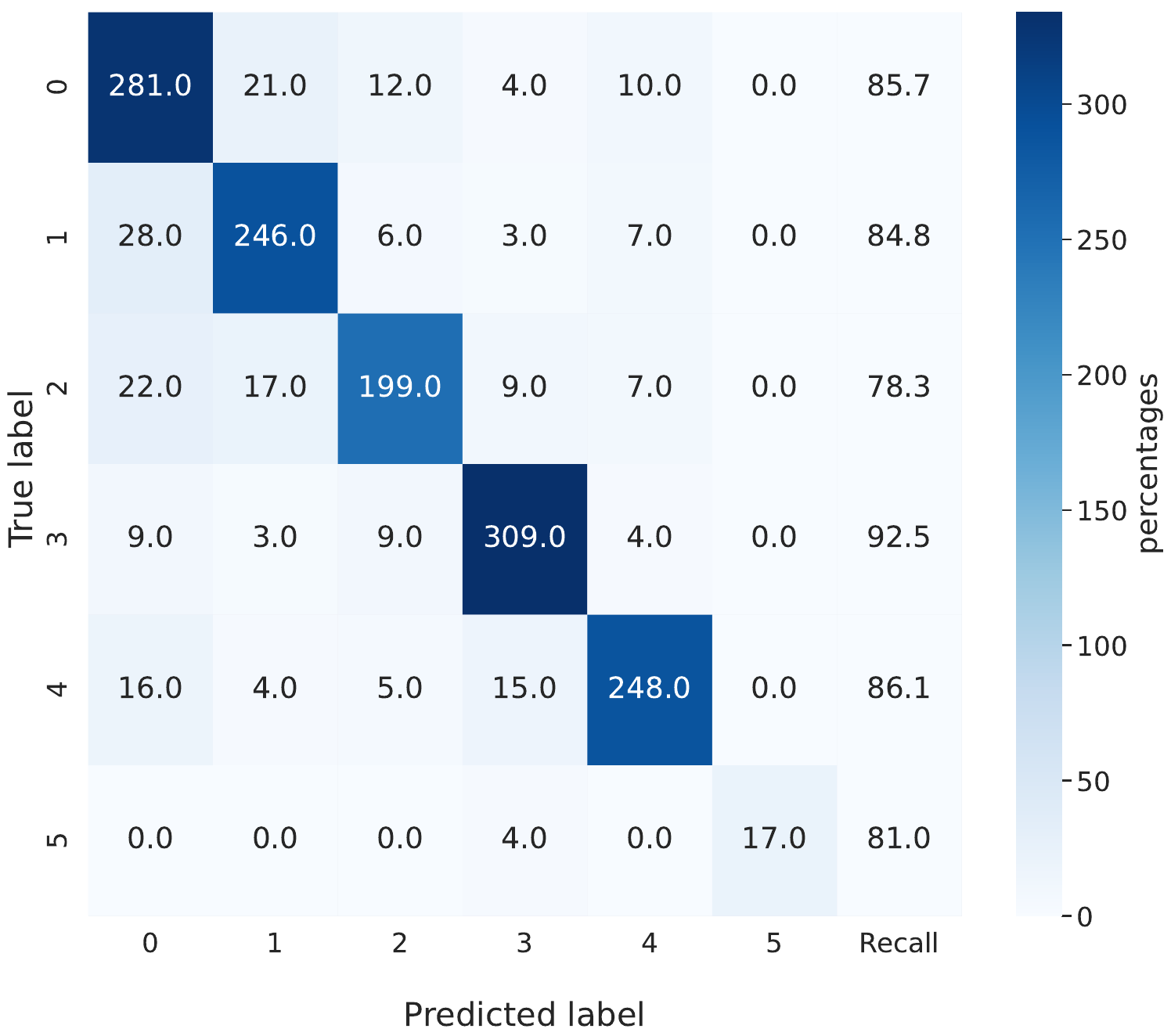}
    \caption{Confusion matrix of six-classes classification. The numbers 0, 1, 2, 3, 4 and 5 indicate different body positions: supine, supine but left lateral head, supine but right lateral head, left-side lying, right-side lying and prone, respectively.}
    \label{confusion}
\end{figure}

Furthermore, the model achieves an accuracy of 85.8\% in six-class classification experiments. The corresponding confusion matrix in Figure \ref{confusion} reveals significant differences between the classes of supine but right lateral head and left-side lying, with a range of 14.2\% in the recall. This disparity may be attributed to data imbalance and limitations in the selected model architecture or model input. Future research should focus on addressing these challenges to enhance the accuracy of algorithms by accounting for the specific differences between the classes.

\section{Conclusions}
\label{conclusion}

The phenomenon of snoring is influenced by the physical position of an individual's body during sleep. Despite this recognition, a lack of comprehensive datasets that consider the role of body position in snoring analysis has been observed. In order to address this shortfall, this paper endeavors to create a snoring dataset called SSBPR that incorporates the aspect of body position, which is expected to provide valuable insights into the field of snoring research. Furthermore, baseline experiments demonstrate that automatic classification models based on acoustic properties can differentiate between snoring at different sleep body positions. Adding more participants to the database, refining the snoring classes, and developing novel methods for snoring sound classification are areas for future work to further enhance the classification performance of different sleep body positions of snoring. The dataset will be released at \url{https://github.com/xiaoli1996/SSBPR}.

\section{Acknowledge}
This work was supported by the Hubei Province Technological Innovation Major Project (No. 2022BCA041).

\newpage

\bibliographystyle{IEEEtran}
\bibliography{mybib}

\end{document}